\documentclass{PoS}
\usepackage{epsfig}
\usepackage{bm}
\usepackage{amssymb}
\usepackage{fontenc}
\usepackage{times}
\usepackage{mathptmx}
\usepackage{graphicx}

\title{QCD with colour-sextet quarks}

\ShortTitle{QCD with colour-sextet quarks}

\author{\speaker{D.~K.~Sinclair}%
         \thanks{This research was supported in part by US Department of Energy
         contracts DE-AC02-06CH11357 and \newline DE-FG02-12ER41871}
         \\
HEP Division, Argonne National Laboratory, 9700 South Cass Avenue, Argonne, 
Illinois 60439, USA\\
and\\
Department of Physics and Astronomy, The University of Iowa, Iowa City, 
Iowa 52242, USA\\
        E-mail: \email{dks@hep.anl.gov}}

\author{J.~B.~Kogut\\
Department of Energy, Division of High Energy Physics, Washington, DC 20585,
USA\\
and\\
Department of Physics -- TQHN, University of Maryland, 82 Regents Drive, 
College Park, MD 20742, USA\\
        E-mail: \email{jbkogut@umd.edu}}

\abstract{
We study QCD with 2 colour-sextet quarks as a model for walking Technicolor,
using lattice gauge theory simulations (RHMC) at finite temperature. Our goal
is to determine if the massless theory is QCD-like (confining, with 
spontaneously-broken chiral symmetry) with a slowly varying coupling (walks) or
if it is a conformal field theory. We do this by simulating the theory at finite
temperature and observing how the coupling at the chiral-symmetry restoration
temperature depends on the temporal extent $N_t$ of the lattice (in lattice 
units). If the theory is QCD-like, this coupling should approach zero in the
large $N_t$ limit in the manner predicted by asymptotic freedom. If it is
conformal, this coupling should approach a finite value in this limit, i.e.
the transition would be a bulk transition. We discuss new results at $N_t=6,8$
and $12$. These preliminary results indicate that the coupling does decrease
with increasing $N_t$, but it is unclear if this is consistent with asymptotic
freedom. 

We also present new results at $N_t=8$ for QCD with 3 colour-sextet quarks.
This theory is believed to be conformal, and is studied for comparison with
the 2-colour case. Preliminary results suggest that $N_t=8$ is still too
small to access the weak coupling limit.
          }

\FullConference{The 30th International Symposium on Lattice Field Theory\\
		 June 24-29,  2012\\
		 Cairns, Australia}

\begin{document}

\section{Introduction}

We are interested in extensions of the standard model with a
composite/strongly-interacting Higgs sector. Technicolor theories 
\cite{Weinberg:1979bn,Susskind:1978ms} -- QCD-like
theories with massless techni-quarks where the techni-pions play the r\^{o}le
of the Higgs field, giving masses to the W and Z -- show the most promise. It
is difficult to suppress flavour-changing neutral currents in extended
Technicolor while giving large enough masses to the fermions. Technicolor
theories which are simply scaled-up QCD fail the precision electroweak tests
\cite{Peskin:1991sw}.
Walking Technicolor theories \cite{Holdom:1981rm,Yamawaki:1985zg,Akiba:1985rr,
Appelquist:1986an},
where the fermion content of the gauge theory is
such that the running coupling constant evolves very slowly over a considerable
range of length/energy-momentum scales, can potentially avoid these problems
\cite{Appelquist:1998xf,Hsu:1998jd,Kurachi:2006mu,Appelquist:2010xv}.

QCD with $1\frac{28}{125} < N_f < 3\frac{3}{10}$ flavours of massless 
colour-sextet quarks is expected to be either a Walking or a Conformal field 
theory. (First term in the $\beta$ function is negative, second positive.)
The $N_f=3$ theory is presumably conformal. 
The $N_f=2$ theory could be either Walking or Conformal. In addition,
if walking, it is minimal, having just the right number of Goldstone bosons (3)
to give masses to the W and Z.
For other arguments as to why this theory might be of interest see
\cite{Sannino:2004qp,Dietrich:2005jn}.)
We simulate the $N_f=2$ theory to see if it walks. We also study the 
$N_f=3$ theory for comparison.
We simulate these theories at finite temperature, using the evolution of
the lattice bare coupling at the chiral transition with $N_t$ to determine
if it is governed by asymptotic freedom -- walking -- or if it approaches a
non-zero constant (bulk transition) -- conformal.
The deconfinement transition occurs at appreciably smaller $\beta=6/g^2$ 
(stronger coupling), and it is unlikely to give useful information on QCD-like 
versus conformal behaviour at the $N_t$ values we use.

We simulate the $N_f=2$ theory on lattices with $N_t=4,6,8,12$ and hope to
extend this to larger $N_t$. (For our earlier work on this theory see
\cite{Kogut:2010cz,Kogut:2011ty,Sinclair:2011ie}
Preliminary results indicate that
$\beta_\chi(N_t=12)$ is significantly larger than $\beta_\chi(N_t=8)$, but by
less than what the 2-loop $\beta$-function would predict. We are also
performing preliminary runs with $\beta$ fixed at a value above
$\beta_\chi(N_t=12)$ on lattices with fixed $N_s$ varying $N_t$, keeping $N_t
\le N_s$ (we thank Julius Kuti for suggesting this). Here we look for the
transition to the chiral-symmetry restored state as $N_t$ increases.

We simulate the $N_f=3$ theory on lattices with $N_t=4,6,8$ and hope to extend
this to $N_t=12$. (For earlier work, see \cite{Kogut:2011bd}.)
Preliminary results indicate that $\beta_\chi(N_t=8)$ is
probably significantly greater than $\beta_\chi(N_t=6)$ which would indicate
that we are not yet at weak enough coupling. 

We are using unimproved staggered quarks and a simple Wilson plaquette action
for all our simulations. The RHMC algorithm is used to implement the required
fractional powers of the fermion determinant.

DeGrand {\it et al.} are studying the 2-flavour theory using improved Wilson
fermions \cite{Shamir:2008pb,DeGrand:2008kx,DeGrand:2009hu,DeGrand:2010na,
DeGrand:2012yq}. Most of their simulations have been at zero temperature.
So far their results are inconclusive, although they tend to favour
the conformal field theory interpretation. The Lattice Higgs Collaboration
are also studying this theory at zero temperature using improved staggered
quarks. \cite{Fodor:2009ar,Fodor:2011tw,Fodor:2012ty,kuti,holland}.
Their results favour a QCD-like (and hence walking) interpretation.

\section{Simulations of lattice QCD with 2 colour-sextet quarks at finite 
temperature}

\subsection{Simulations at $\bm{N_t=6}$}

We have extended our simulations for quark mass $m=0.005$ on a $12^3 \times 6$
lattice, such that the $\beta$ spacing is now $0.02$ in the range 
$6.5 \le \beta \le 6.7$, which encompasses the chiral transition. In this range
we have accumulated 100,000 trajectories for each $\beta$.

At this quark mass, it is impossible to determine the position of the 
chiral-symmetry restoration transition with any precision from the chiral
condensate itself. We determine the position of this transition from the peak
in the disconnected chiral susceptibility. This is defined by:
\begin{equation}
\chi_{\bar{\psi}\psi} = \frac{V}{T}\left[\langle (\bar{\psi}\psi)^2 \rangle
                                   -(\langle \bar{\psi}\psi \rangle)^2\right]
\end{equation}
where $V$ is the spatial volume of the lattice and $T$ is the temperature.
$\bar{\psi}\psi$ is a lattice averaged quantity. Because we only have 
stochastic estimators for $\bar{\psi}\psi$ (5 per trajectory), we obtain 
unbiased estimators of $(\bar{\psi}\psi)^2$ as the products of 2 different
estimators of $\bar{\psi}\psi$ for the same gauge configuration.

\begin{figure}[htb]
\parbox{2.9in}{
\epsfxsize=2.9in
\centerline{\epsffile{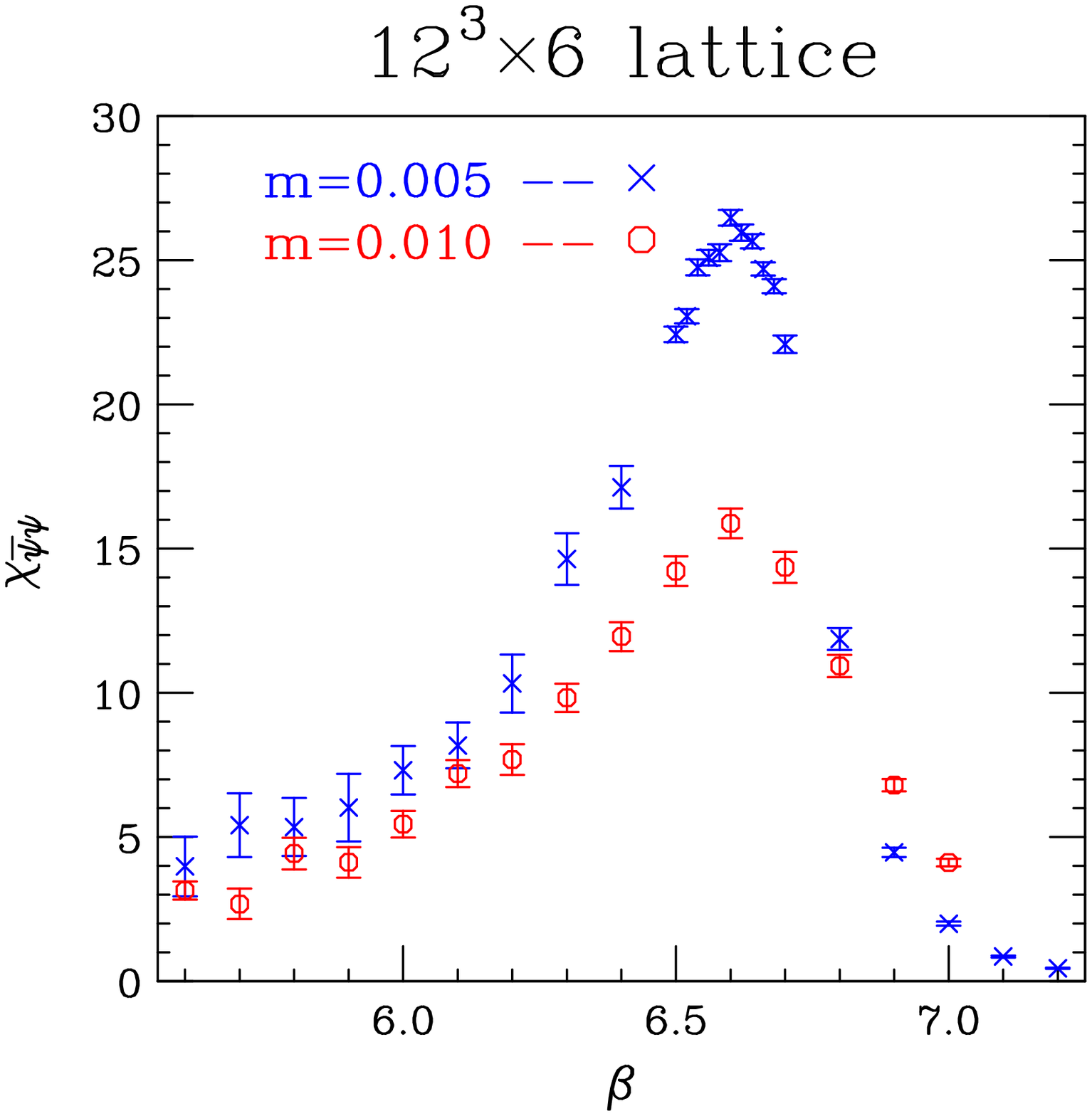}}
\caption{Chiral susceptibilities on a $12^3 \times 6$ lattice, $N_f=2$.}
\label{fig:chi6}
}
\parbox{0.2in}{}
\parbox{2.9in}{
\epsfxsize=2.9in                                                               
\centerline{\epsffile{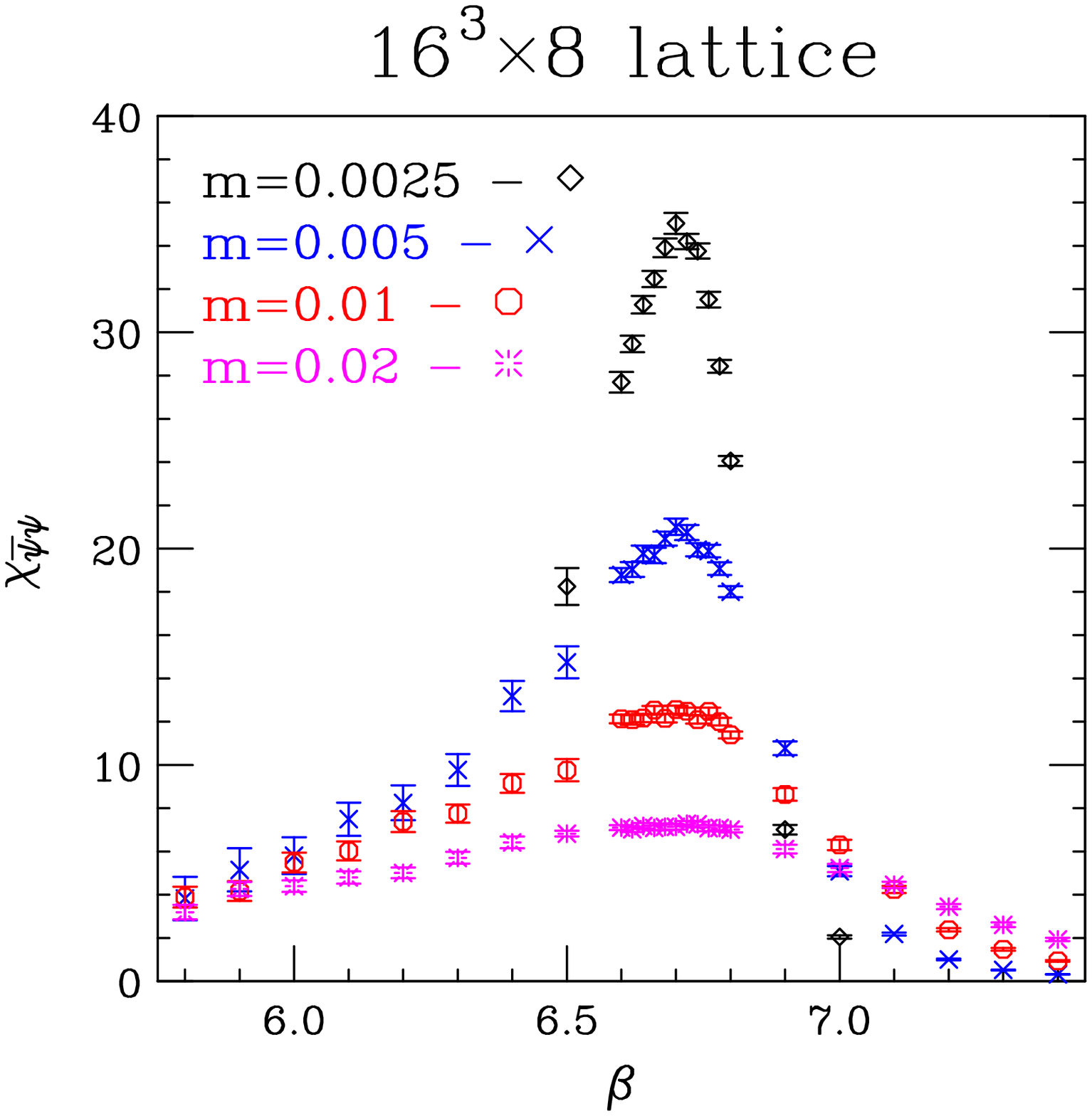}}                              
\caption{Chiral susceptibilities on a $16^3 \times 8$ lattice, $N_f=2$.}
\label{fig:chi8}
}
\end{figure}

Figure~\ref{fig:chi6} shows these chiral susceptibilities for our 
$12^3 \times 6$ simulations. The $m=0.005$ susceptibility shows a clear peak
at $\beta \approx 6.60$. Under the assumption that the position of this peak
has little mass dependence (which is not inconsistent with this data), we
thus conclude that the position of the chiral transition in the chiral 
($m \rightarrow 0$) limit is at $\beta=\beta_\chi=6.60(2)$

\subsection{Simulations at $\bm{N_t=8}$}

We have performed simulations on $16^3 \times 8$ lattices in the vicinity of
the chiral transition. Quark masses $m=0.0025$, $m=0.005$, $m=0.01$ and 
$m=0.02$ were used, to enable extrapolation to the chiral limit. For 
$6.6 \le \beta \le 6.8$ which spans the neighbourhood of the chiral transition
we performed simulations at $\beta$ values spaced by $0.02$ to enable accurate
determination of the position of this transition. At $m=0.0025$ we performed
simulations of 100,000 trajectories in length for each $\beta$ in this range.
For each of the 3 larger masses we simulated 50,000 trajectories at each of
the $\beta$ values in this range.

Again, the dependence of the chiral condensates on $\beta$ is too smooth to
accurately determine the position of the chiral transition, even though the
decrease in the chiral condensate with decreasing mass does indicate that there
{\it is} such a transition above which the condensate vanishes in the chiral
limit. Thus we again turn to the chiral susceptibility, which has a peak that
diverges in the chiral limit, to accurately determine $\beta_\chi$. A graph of
these chiral susceptibilities for each of the 4 masses is shown in 
figure~\ref{fig:chi8}.

Because the $\beta$ values used are sufficiently close together, it is possible
to use Ferrenberg-Swendsen reweighting to determine the position of the peak
for $m=0.0025$ more precisely. This yields $\beta=\beta_{\it peak}=6.690(5)$.
Since $\beta_{\it peak}$ clearly has very little mass dependence, we deduce
that $\beta_\chi=6.69(1)$. Hence
\begin{equation}
\beta_\chi(N_t=8)-\beta_\chi(N_t=6) \approx 0.09 ,
\end{equation}
consistent with the 2-loop perturbative prediction of $\approx 0.087$.

\subsection{Simulations at $\bm{N_t=12}$}

We are simulating lattice QCD with 2 flavours of light colour-sextet quarks
on a $24^3 \times 12$ lattice, in the neighbourhood of the chiral-symmetry
restoration transition. We are running at masses $m=0.0025$, $m=0.005$ and
$m=0.01$, to enable continuation to the chiral limit. For 
$6.6 \le \beta \le 6.9$, we are running at a set of $\beta$s spaced by $0.02$.
In this range, we have 10,000 -- 25,000 trajectories at each $\beta$ and $m$
(except at $\beta=6.6$, $m=0.005$, where we have 62,500 trajectories), and
will extend this to 50,000 -- 100,000 trajectories.

\begin{figure}[htb]
\parbox{2.9in}{
\epsfxsize=2.9in
\centerline{\epsffile{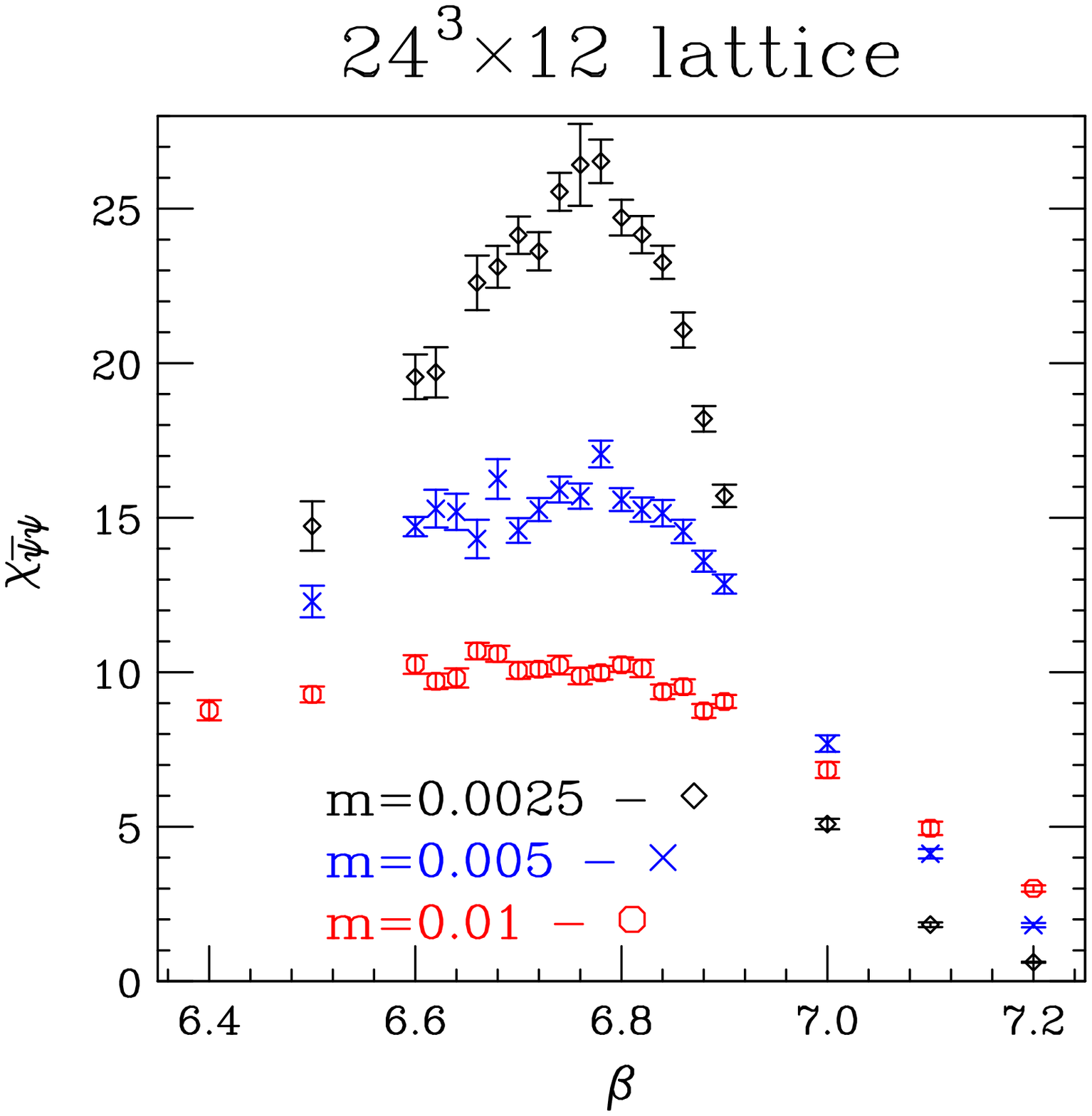}}
\caption{Chiral susceptibilities on a $24^3 \times 12$ lattice, $N_f=2$.}
\label{fig:chi12}
}
\parbox{0.2in}{}
\parbox{2.9in}{
\epsfxsize=2.9in
\centerline{\epsffile{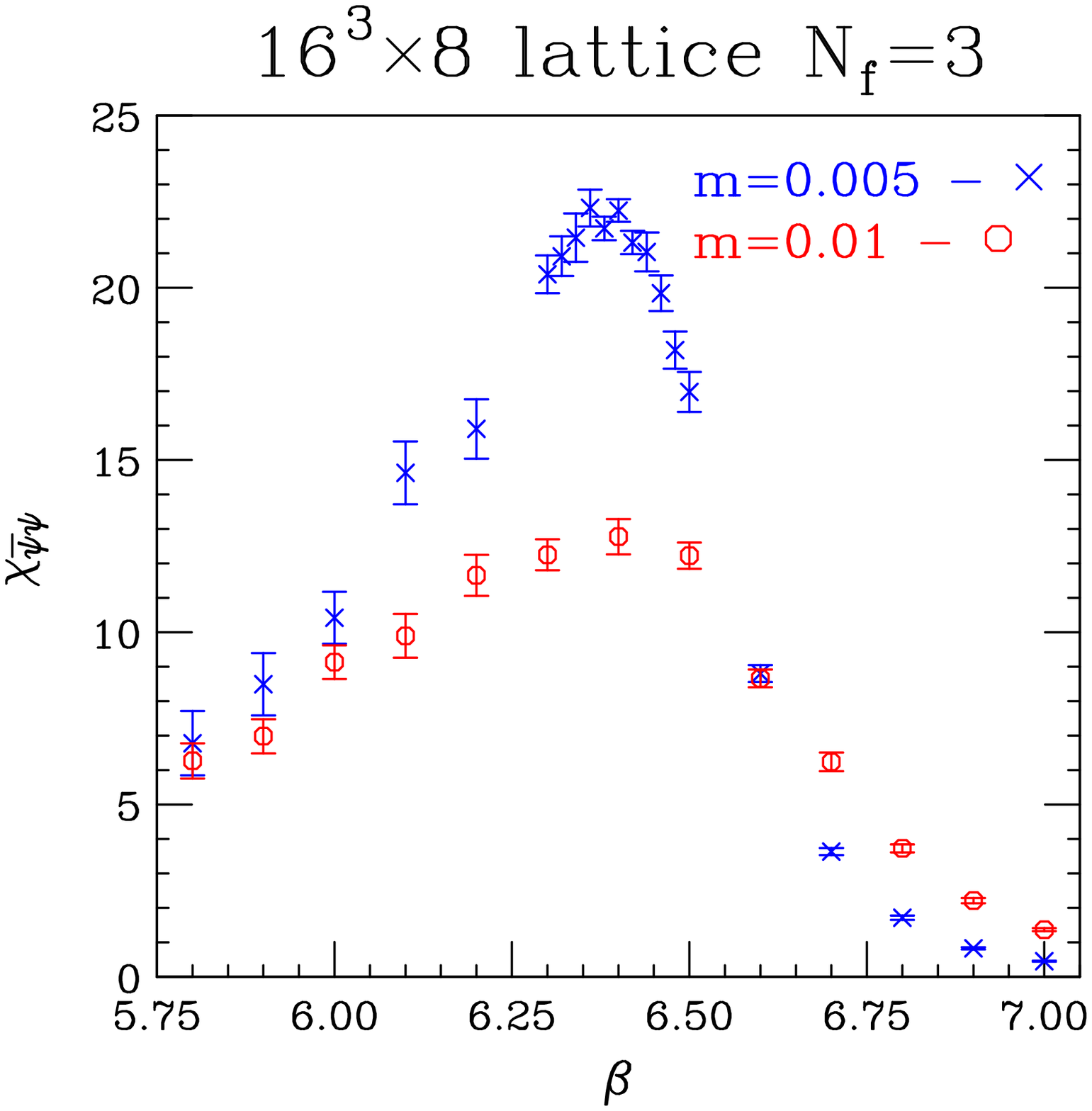}}
\caption{Chiral susceptibilities on a $16^3 \times 8$ lattice, $N_f=3$.}
\label{fig:chi3_8}
}
\end{figure}

Figure~\ref{fig:chi12} shows the chiral susceptibilities from these runs.
While the $m=0.0025$ and possibly the $m=0.005$ susceptibilities show clear
peaks, the statistics are not yet adequate to make a precise estimate of the
positions of these peaks. The $m=0.0025$ graph shows a peak at 
$\beta=\beta_\chi \approx 6.78$. If this were correct, it would mean that
$\beta_\chi(N_t=12)-\beta_\chi(N_t=8) \approx 0.09$, compared with the predicted
$0.12$.

\subsection{Other simulations and analyses}

In addition to measuring the unsubtracted, unrenormalized chiral condensates,
we have also used the subtraction scheme of the Lattice Higgs Collaboration,
which removes much of the quadratic UV divergence.
\begin{equation}
\langle{\bar{\psi}\psi}\rangle_{sub} = \langle{\bar{\psi}\psi}\rangle 
-\left(m_V\frac{\partial}{\partial m_V}\langle{\bar{\psi}\psi}\rangle\right)
                                                                  _{m_V=m}
\end{equation}
\cite{Fodor:2011tw},
where $m_V$ is the valence-quark mass. While this reduces the magnitude of
$\langle{\bar{\psi}\psi}\rangle$, making the fact that it vanishes at large
$\beta$ more obvious, it does not help significantly in the precision 
determination of $\beta_\chi$ at the masses we use.

Finally, we are simulating our theory on $24^3 \times N_t$ lattices for
$N_t \le 24$ at $\beta=6.9$ to try to observe chiral-symmetry restoration as
$N_t$ is increased. Here we look at both the unsubtracted and the subtracted
chiral condensates. Our preliminary results look promising but not conclusive. 

\section{Simulations of lattice QCD with 3 colour-sextet quarks at finite
temperature}

We are extending our earlier simulations of QCD with 3 colour-sextet quarks
to $N_t=8$. Our runs have been performed on $16^3 \times 8$ lattices with 
$m=0.005$ and $m=0.01$. For $m=0.005$ and $6.3 \le \beta \le 6.5$ (the 
neighbourhood of the chiral transition), we currently have 20,000 to 50,000
trajectories for each $\beta$. Elsewhere we have 10,000 trajectories for each
$(\beta,m)$.

Figure~\ref{fig:chi3_8} shows the chiral susceptibilities from these runs.
While we need to complete 50,000 trajectories at each $\beta$ in the above
range to accurately determine $\beta_\chi$, it already appears that \newline 
$\beta_\chi(N_t=8)-\beta_\chi(N_t=6)$ will be appreciable. Since we believe
that this theory is conformal, when $\beta_\chi(N_t)$ will approach a finite
value as $N_t \rightarrow \infty$, this suggests that we will need a larger 
$N_t$ to see the weak-coupling limit. Even if this theory were QCD-like,
the 2-loop $\beta$-function would predict 
$\beta_\chi(N_t=8)-\beta_\chi(N_t=6) \approx 0.0025$, which would be too small
for us to see, and hence inconsistent with our preliminary observations.

\section{Discussion and Conclusions}

We simulate the thermodynamics of QCD with 2 colour-sextet quarks on lattices
with \newline $N_t=4,6,8,12$, in the neighbourhood of the chiral transition. If
chiral-symmetry restoration is a finite-temperature phase transition, measuring
$\beta_\chi$ as a function of $N_t$ yields the running of the bare lattice
coupling $\beta$ with $a$ at $a=1/(N_tT_\chi)$. Asymptotic freedom would imply
that $\beta_\chi \rightarrow \infty$ ($g_\chi^2 \rightarrow 0$) as 
$N_t \rightarrow \infty$ and hence $a \rightarrow 0$. At the masses we use,
the unsubtracted, unrenormalized chiral condensates become small at large
$\beta$s, and appear consistent with extrapolating to zero as $m \rightarrow
0$. However, they decrease too smoothly to enable an accurate determination of
$\beta_\chi$. Hence we have extracted $\beta_\chi$ from the peaks in the
chiral susceptibilities. We have also examined the chiral condensates with the
subtraction scheme used by the Lattice Higgs Collaboration. While these
subtracted condensates show more clearly that the condensates will vanish in
the chiral limit at large enough $\beta$, they still do not allow for an
accurate determination of $\beta_\chi$.

We present preliminary results indicating that $\beta_\chi$ increases with
$N_t$ over the range of $N_t$s considered. This suggests that the theory does
walk. The change in $\beta_\chi$ between $N_t=6$ and $N_t=8$ is in excellent
agreement with 2-loop perturbation theory. However, the change in $\beta_\chi$
between $N_t=8$ and $N_t=12$ appears to be about 25\% smaller than would be
predicted from the 2-loop $\beta$-function. This is of concern, since this
lies in the assumed weak-coupling domain ($\beta \stackrel{>}{_\sim}
\beta_\chi(N_t=6)$). However, we need more statistics to accurately determine
$\beta_\chi(N_t=12)$, and if it remains low, we will need to check that this is
not a finite volume effect. It is possible that the 2-loop $\beta$-function is
inadequate to describe the running of the bare coupling for unimproved
staggered lattice QCD at these couplings. This will then require simulations
at larger $N_t$, to distinguish walking from conformal behaviour.

A series of runs performed on $24^3 \times N_t$ lattices for several $N_t$s
($N_t \le 24$) at small quark masses, and a fixed coupling intermediate between
$\beta_\chi(N_t=12)$ and the expected value of $\beta_\chi(N_t=24)$, does show
an increase in both the unsubtracted and subtracted chiral condensates. More
work is needed to determine if this implies a transition to a chirally broken
theory as $N_t$ increases.

The zero temperature properties of this theory need to be studied and the
results compared with Fodor {\it et al} and DeGrand {\it et al}. In
particular we need to determine the masses of the Higgs and the dilaton, to
see if either of these could be the Higgs-like particle observed at the LHC.
For a recent discussion of what this entails see reference~\cite{Fodor:2012ty}.

We are extending our $N_f=3$ runs to $N_t=8$. Preliminary results
indicate that there is a substantial increase in $\beta_\chi$ between
$N_t=6$ and $N_t=8$. This would indicate that, for this range of $N_t$s, 
$\beta_\chi$ does not lie completely in the weak-coupling domain, and that we 
will need simulations at larger $N_t$. 

Other theories we plan to study include $SU(2)_{colour}$ with 3 
colour-adjoint (symmetric) Majorana/Weyl quarks, and $SU(4)_{colour}$ with 
colour-antisymmetric quarks. 

\section*{Acknowledgements}

This research used resources of the National Energy Research Scientific
Computing Center, which is supported by the Office of Science of the U.S.
Department of Energy under Contract No. DE-AC02-05CH11231. In particular,
these simulations were performed on the Cray XT4, Franklin, the Cray XT6,
Hopper, and the Linux cluster, Carver, all at NERSC. In addition this research
used the Cray XT5, Kraken at NICS under XSEDE Project Number: TG-MCA99S015. We
also acknowledge the computing resources provided on ``Fusion'', a 320-node
computing cluster operated by the Laboratory Computing Resource Center at
Argonne National Laboratory.


\begin{thebibliography}{99}
%% Technicolor

%\cite{Weinberg:1979bn}
\bibitem{Weinberg:1979bn}
  S.~Weinberg,
  %``Implications Of Dynamical Symmetry Breaking: An Addendum,''
  Phys.\ Rev.\  D {\bf 19}, 1277 (1979).
  %%CITATION = PHRVA,D19,1277;%%

%\cite{Susskind:1978ms}
\bibitem{Susskind:1978ms}
  L.~Susskind,
  %``Dynamics Of Spontaneous Symmetry Breaking In The Weinberg-Salam Theory,''
  Phys.\ Rev.\  D {\bf 20}, 2619 (1979).
  %%CITATION = PHRVA,D20,2619;%%

%% Precision Electroweak.

%\cite{Peskin:1990zt}
\bibitem{Peskin:1990zt} 
  M.~E.~Peskin and T.~Takeuchi,
  %``A New constraint on a strongly interacting Higgs sector,''
  Phys.\ Rev.\ Lett.\  {\bf 65}, 964 (1990).
  %%CITATION = PRLTA,65,964;%%

%\cite{Peskin:1991sw}
\bibitem{Peskin:1991sw} 
  M.~E.~Peskin and T.~Takeuchi,
  %``Estimation of oblique electroweak corrections,''
  Phys.\ Rev.\ D {\bf 46}, 381 (1992).
  %%CITATION = PHRVA,D46,381;%%

%% Walking Technicolor

%\cite{Holdom:1981rm}
\bibitem{Holdom:1981rm}
  B.~Holdom,
  %``Raising The Sideways Scale,''
  Phys.\ Rev.\  D {\bf 24}, 1441 (1981).
  %%CITATION = PHRVA,D24,1441;%%

%\cite{Yamawaki:1985zg}
\bibitem{Yamawaki:1985zg}
  K.~Yamawaki, M.~Bando and K.~i.~Matumoto,
  %``Scale Invariant Technicolor Model And A Technidilaton,''
  Phys.\ Rev.\ Lett.\  {\bf 56}, 1335 (1986).
  %%CITATION = PRLTA,56,1335;%%

%\cite{Akiba:1985rr}
\bibitem{Akiba:1985rr}
  T.~Akiba and T.~Yanagida,
  %``Hierarchic Chiral Condensate,''
  Phys.\ Lett.\  B {\bf 169}, 432 (1986).
  %%CITATION = PHLTA,B169,432;%%

%\cite{Appelquist:1986an}
\bibitem{Appelquist:1986an}
  T.~W.~Appelquist, D.~Karabali and L.~C.~R.~Wijewardhana,
  %``Chiral Hierarchies and the Flavor Changing Neutral Current Problem in
  %Technicolor,''
  Phys.\ Rev.\ Lett.\  {\bf 57}, 957 (1986).
  %%CITATION = PRLTA,57,957;%%

%% Precision electroweak and (walking) Technicolor

%\cite{Appelquist:1998xf}
\bibitem{Appelquist:1998xf} 
  T.~Appelquist and F.~Sannino,
  %``The Physical spectrum of conformal SU(N) gauge theories,''
  Phys.\ Rev.\ D {\bf 59}, 067702 (1999)
  [hep-ph/9806409].
  %%CITATION = HEP-PH/9806409;%%

%\cite{Hsu:1998jd}
\bibitem{Hsu:1998jd} 
  S.~D.~H.~Hsu, F.~Sannino and J.~Schechter,
  %``Anomaly induced QCD potential and quark decoupling,''
  Phys.\ Lett.\ B {\bf 427}, 300 (1998)
  [hep-th/9801097].
  %%CITATION = HEP-TH/9801097;%%

%\cite{Kurachi:2006mu}
\bibitem{Kurachi:2006mu} 
  M.~Kurachi and R.~Shrock,
  %``Behavior of the S Parameter in the Crossover Region Between Walking and 
  % QCD-Like Regimes of an SU(N) Gauge Theory,''
  Phys.\ Rev.\ D {\bf 74}, 056003 (2006)
  [hep-ph/0607231].
  %%CITATION = HEP-PH/0607231;%%

%\cite{Appelquist:2010xv}
\bibitem{Appelquist:2010xv} 
  T.~Appelquist {\it et al.}  [LSD Collaboration],
  %``Parity Doubling and the S Parameter Below the Conformal Window,''
  Phys.\ Rev.\ Lett.\  {\bf 106}, 231601 (2011)
  [arXiv:1009.5967 [hep-ph]].
  %%CITATION = ARXIV:1009.5967;%%

%% Justification for QCD with N_f=2 sextet quarks.

%\cite{Sannino:2004qp}
\bibitem{Sannino:2004qp}
  F.~Sannino, K.~Tuominen,
  %``Orientifold theory dynamics and symmetry breaking,''
  Phys.\ Rev.\  {\bf D71}, 051901 (2005).
  [hep-ph/0405209].

%\cite{Dietrich:2005jn}
\bibitem{Dietrich:2005jn}
  D.~D.~Dietrich, F.~Sannino, K.~Tuominen,
  %``Light composite Higgs from higher representations versus electroweak pr
% ecision measurements: Predictions for CERN LHC,''
  Phys.\ Rev.\  {\bf D72}, 055001 (2005).
  [hep-ph/0505059].

%% Our work on N_f=2 sextet quarks

%\cite{Kogut:2010cz}
\bibitem{Kogut:2010cz}
  J.~B.~Kogut, D.~K.~Sinclair,
  %``Thermodynamics of lattice QCD with 2 flavours of colour-sextet quarks: A 
  %model of walking/conformal Technicolor,''
  Phys.\ Rev.\  {\bf D81}, 114507 (2010).
  [arXiv:1002.2988 [hep-lat]].

%\cite{Kogut:2011ty}
\bibitem{Kogut:2011ty} 
  J.~B.~Kogut and D.~K.~Sinclair,
  %``Thermodynamics of lattice QCD with 2 sextet quarks on $N_t$=8 lattices,''
  Phys.\ Rev.\ D {\bf 84}, 074504 (2011)
  [arXiv:1105.3749 [hep-lat]].
  %%CITATION = ARXIV:1105.3749;%%

%\cite{Sinclair:2011ie}
\bibitem{Sinclair:2011ie} 
  D.~K.~Sinclair and J.~B.~Kogut,
  %``The chiral phase transition for QCD with sextet quarks,''
  PoS LATTICE {\bf 2011}, 090 (2011)
  [arXiv:1111.2319 [hep-lat]].
  %%CITATION = ARXIV:1111.2319;%%

%% Our work on N_f=3 sextet quarks

%\cite{Kogut:2011bd}
\bibitem{Kogut:2011bd} 
  J.~B.~Kogut and D.~K.~Sinclair,
  %``Thermodynamics of lattice QCD with 3 flavours of colour-sextet quarks,''
  Phys.\ Rev.\ D {\bf 85}, 054505 (2012)
  [arXiv:1111.3353 [hep-lat]].
  %%CITATION = ARXIV:1111.3353;%%

%% Degrand et al on QCD with colour sextet quarks

%\cite{Shamir:2008pb}
\bibitem{Shamir:2008pb} 
  Y.~Shamir, B.~Svetitsky and T.~DeGrand,
  %``Zero of the discrete beta function in SU(3) lattice gauge theory with color sextet fermions,''
  Phys.\ Rev.\ D {\bf 78}, 031502 (2008)
  [arXiv:0803.1707 [hep-lat]].
  %%CITATION = ARXIV:0803.1707;%%

%\cite{DeGrand:2008kx}
\bibitem{DeGrand:2008kx} 
  T.~DeGrand, Y.~Shamir and B.~Svetitsky,
  %``Phase structure of SU(3) gauge theory with two flavors of 
  % symmetric-representation fermions,''
  Phys.\ Rev.\ D {\bf 79}, 034501 (2009)
  [arXiv:0812.1427 [hep-lat]].
  %%CITATION = ARXIV:0812.1427;%%

%\cite{DeGrand:2009hu}
\bibitem{DeGrand:2009hu} 
  T.~DeGrand,
  %``Finite-size scaling tests for SU(3) lattice gauge theory with color sextet
  % fermions,''
  Phys.\ Rev.\ D {\bf 80}, 114507 (2009)
  [arXiv:0910.3072 [hep-lat]].
  %%CITATION = ARXIV:0910.3072;%%

%\cite{DeGrand:2010na}
\bibitem{DeGrand:2010na} 
  T.~DeGrand, Y.~Shamir and B.~Svetitsky,
  %``Running coupling and mass anomalous dimension of SU(3) gauge theory with 
  % two flavors of symmetric-representation fermions,''
  Phys.\ Rev.\ D {\bf 82}, 054503 (2010)
  [arXiv:1006.0707 [hep-lat]].
  %%CITATION = ARXIV:1006.0707;%%

%\cite{DeGrand:2012yq}
\bibitem{DeGrand:2012yq} 
  T.~DeGrand, Y.~Shamir and B.~Svetitsky,
  %``Mass anomalous dimension in sextet QCD,''
  arXiv:1201.0935 [hep-lat].
  %%CITATION = ARXIV:1201.0935;%%

%% Lattice Higgs Collaboration on QCD with colour sextet quarks

%\cite{Fodor:2009ar}
\bibitem{Fodor:2009ar} 
  Z.~Fodor, K.~Holland, J.~Kuti, D.~Nogradi and C.~Schroeder,
  %``Chiral properties of SU(3) sextet fermions,''
  JHEP {\bf 0911}, 103 (2009)
  [arXiv:0908.2466 [hep-lat]].
  %%CITATION = ARXIV:0908.2466;%%

%\cite{Fodor:2011tw}
\bibitem{Fodor:2011tw} 
  Z.~Fodor, K.~Holland, J.~Kuti, D.~Nogradi and C.~Schroeder,
  %``Chiral symmetry breaking in fundamental and sextet fermion representations 
  % of SU(3) color,''
  arXiv:1103.5998 [hep-lat].
  %%CITATION = ARXIV:1103.5998;%%

%\cite{Fodor:2012ty}
\bibitem{Fodor:2012ty} 
  Z.~Fodor, K.~Holland, J.~Kuti, D.~Nogradi, C.~Schroeder and C.~H.~Wong,
  %``Can the nearly conformal sextet gauge model hide the Higgs impostor?,''
  arXiv:1209.0391 [hep-lat].
  %%CITATION = ARXIV:1209.0391;%%

%\cite{kuti}
\bibitem{kuti}
  J.~Kuti, talk presented at Lattice 2012, Cairns, Australia.

%\cite{holland}
\bibitem{holland}
  K.~Holland, talk presented at Lattice 2012, Cairns, Australia.

\end{thebibliography}
\end{document}